\documentclass[twocolumn,twocolappendix]{aastex631}
\usepackage{amsmath}

\received{August 8, 2023}
\revised{January 18, 2024}
\accepted{January 27, 2024}

\shorttitle{}
\shortauthors{Yamada et al.}
\graphicspath{{./}{figures/}}

\begin{document}

\title{[\ion{O}{4}] and [\ion{Ne}{5}]-weak AGNs Hidden by Compton-thick Material in Late Mergers}

\author[0000-0002-9754-3081]{Satoshi Yamada}
\affiliation{RIKEN Cluster for Pioneering Research, 2-1 Hirosawa, Wako, Saitama 351-0198, Japan; satoshi.yamada@riken.jp}

\author[0000-0001-7821-6715]{Yoshihiro Ueda}
\affiliation{Department of Astronomy, Kyoto University, Kitashirakawa-Oiwake-cho, Sakyo-ku, Kyoto 606-8502, Japan}

\author[0000-0002-6808-2052]{Taiki Kawamuro}
\affiliation{RIKEN Cluster for Pioneering Research, 2-1 Hirosawa, Wako, Saitama 351-0198, Japan; satoshi.yamada@riken.jp}

\author[0000-0001-5231-2645]{Claudio Ricci}
\affiliation{N\'ucleo de Astronom\'{\i}a de la Facultad de Ingenier\'{\i}a, Universidad Diego Portales, Av. Ej\'ercito Libertador 441, Santiago, Chile}
\affiliation{Kavli Institute for Astronomy and Astrophysics, Peking University, Beijing 100871, People's Republic of China}
\affiliation{George Mason University, Department of Physics \& Astronomy, MS 3F3, 4400 University Drive, Fairfax, VA 22030, USA}

\author[0000-0002-3531-7863]{Yoshiki Toba}
\affiliation{National Astronomical Observatory of Japan, 2-21-1 Osawa, Mitaka, Tokyo 181-8588, Japan}
\affiliation{Department of Physics, Nara Women’s University, Kitauoyanishi-machi, Nara, Nara
630-8506, Japan}
\affiliation{Academia Sinica Institute of Astronomy and Astrophysics, 11F of Astronomy-Mathematics
Building, AS/NTU, No.1, Section 4, Roosevelt Road, Taipei 10617, Taiwan}
\affiliation{Research Center for Space and Cosmic Evolution, Ehime University, 2-5 Bunkyo-cho, Matsuyama, Ehime 790-8577, Japan}

\author[0000-0001-6186-8792]{Masatoshi Imanishi}
\affiliation{National Astronomical Observatory of Japan, 2-21-1 Osawa, Mitaka, Tokyo 181-8588, Japan}
\affiliation{Department of Astronomical Science, Graduate University for Advanced Studies (SOKENDAI), 2-21-1 Osawa, Mitaka, Tokyo 181-8588, Japan}

\author[0000-0002-7562-485X]{Takamitsu Miyaji}
\affiliation{Instituto de Astronom\'{\i}a sede Ensenada, Universidad Nacional Aut\'onoma de M\'exico, Km 107, Carret. Tij.-Ens., Ensenada, 22060, BC, M\'exico}

\author[0000-0002-0114-5581]{Atsushi Tanimoto}
\affiliation{Graduate School of Science and Engineering, Kagoshima University, Kagoshima 890-0065, Japan}

\author[0000-0002-4377-903X]{Kohei Ichikawa}
\affiliation{Global Center for Science and Engineering, Faculty of
Science and Engineering, Waseda University}
\affiliation{Department of Physics, School of Advanced Science and
Engineering, Faculty of Science and Engineering, Waseda University,
3-4-1,
Okubo, Shinjuku, Tokyo 169-8555, Japan}

\author[0000-0002-8653-020X]{Mart\'{\i}n Herrera-Endoqui}
\affiliation{Instituto de Astronom\'{\i}a sede Ensenada, Universidad Nacional Aut\'onoma de M\'exico, Km 107, Carret. Tij.-Ens., Ensenada, 22060, BC, M\'exico}

\author[0000-0002-5701-0811]{Shoji Ogawa}
\affiliation{Institute of Space and Astronautical Science (ISAS), Japan Aerospace Exploration Agency (JAXA), 3-1-1 Yoshinodai, Chuo-ku, Sagamihara, Kanagawa
252-5210, Japan}

\author[0000-0001-6653-779X]{Ryosuke Uematsu}
\affiliation{Department of Astronomy, Kyoto University, Kitashirakawa-Oiwake-cho, Sakyo-ku, Kyoto 606-8502, Japan}

\author[0000-0002-8779-8486]{Keiichi Wada}
\affiliation{Graduate School of Science and Engineering, Kagoshima University, Kagoshima 890-0065, Japan}
\affiliation{Research Center for Space and Cosmic Evolution, Ehime University, Matsuyama 790-8577, Japan}
\affiliation{Faculty of Science, Hokkaido University, Sapporo 060-0810, Japan}



\begin{abstract}
We study ``buried'' active galactic nuclei (AGNs) almost fully covered by circumnuclear material in ultra-/luminous infrared galaxies (U/LIRGs), which show weak ionized lines from narrow line regions.
\textcolor{black}{Employing an indicator of [\ion{O}{4}] 25.89-$\mu$m or [\ion{Ne}{5}] 14.32-$\mu$m line to 12-$\mu$m AGN luminosity ratio, we find 17 buried AGN candidates that are [\ion{O}{4}]-weak ($L_{\rm [O\,IV]}$/$L_{\rm 12,AGN} \leq -$3.0) or [\ion{Ne}{5}]-weak ($L_{\rm [Ne\,V]}$/$L_{\rm 12,AGN} \leq -$3.4) among 30 AGNs in local U/LIRGs.}
For the [\ion{O}{4}]-weak AGNs, we estimate their covering fractions of Compton-thick (CT; $N_{\rm H} \geq 10^{24}$~cm$^{-2}$) material with an X-ray clumpy torus model to be $f^{\rm (spec)}_{\rm CT} = 0.55\pm0.19$ on average.
This value is consistent with the fraction of CT AGNs ($f^{\rm (stat)}_{\rm CT} = 53\pm12$\%) among the [\ion{O}{4}]-weak AGNs in U/LIRGs and much larger than that in Swift/BAT AGNs ($23\pm6$\%).
The fraction of [\ion{O}{4}]-weak AGNs increases from $27^{+13}_{-10}$\% (early) to $66^{+10}_{-12}$\% (late mergers).
Similar results are obtained with the [\ion{Ne}{5}] line.
The [\ion{O}{4}] or [\ion{Ne}{5}]-weak AGNs in late mergers show larger $N_{\rm H}$ and Eddington ratios ($\lambda_{\rm Edd}$) than those of the Swift/BAT AGNs, and the largest $N_{\rm H}$ is $\gtrsim$10$^{25}$~cm$^{-2}$ at ${\log}\lambda_{\rm Edd} \sim -$1, close to the effective Eddington limit for CT material.
These suggest that (1) the circumnuclear material in buried AGNs is regulated by the radiation force from high-$\lambda_{\rm Edd}$ AGNs on the CT obscurers, and 
(2) their dense material with large $f^{\rm (spec)}_{\rm CT}$ ($\sim 0.5 \pm 0.1$) in U/LIRGs is a likely cause of a unique structure of buried AGNs, whose amount of material may be maintained through merger-induced supply from their host galaxies.

\end{abstract}




\section{Introduction}
Luminous infrared (IR) galaxies (LIRG; with 8--1000~$\mu$m luminosities of $L_{\rm IR} \geq 10^{11} L_{\odot}$) and ultraluminous IR galaxies (ULIRG; $L_{\rm IR} \geq 10^{12} L_{\odot}$) are mostly gas-rich mergers of galaxies \citep[e.g.,][]{Kartaltepe2010}, 
and hence are ideal laboratories to examine the evolution of supermassive black holes (SMBHs) induced by galaxy mergers.
Although they are rare in the local universe, they are the standard population constituting most of the cosmic IR background at $z \gtrsim 1$ \citep[e.g.,][]{Casey2014,Coppin2015}.
Theoretical studies \citep[e.g.,][]{Hopkins2006,Hopkins2008,Blecha2018,Yutani2022} suggest that mergers enhance starburst and mass accretion onto SMBHs, identified as active galactic nuclei (AGNs).
Indeed, multiwavelength AGN surveys have found that AGNs with moderate bolometric luminosities are usually found in nonmergers, whereas more luminous AGNs are commonly found in mergers  \citep[e.g.,][]{Treister2012}.

\label{S1-intro}

The environment in the vicinity of the accreting SMBH is well understood for nonmergers, owing to a large sample size available in the local universe.
A key component for SMBH growth is circumnuclear material, which is fed to the central black hole. 
\citet{Ricci2017cNature} made a systematic study of X-ray luminous AGNs detected with Swift/BAT, where the majority of the hosts are isolated galaxies,
and found that about 70\% of them are obscured by gas and dust with a hydrogen column density of $N_{\rm H} \geq 10^{22}$~cm$^{-2}$.
According to a simple, orientation-based AGN unification model \citep[e.g.,][]{Antonucci1993}, 
\textcolor{black}{if the survey has a high completeness level, the statistical fraction of obscured AGNs, $f_{\rm obs}^{\rm (stat)}$($\geq$10$^{22}$~cm$^{-2}$), should be roughly equivalent to the ``covering fraction'' of the AGN.}
\citet{Ricci2017cNature} also found that the fraction of obscured AGNs in their sample is $f_{\rm obs}^{\rm (stat)} \sim 30\%$ at high Eddington ratios of ${\log}\lambda_{\rm Edd} \gtrsim -$2.
\textcolor{black}{This value of $\lambda_{\rm Edd}$ corresponds to the threshold above which radiation pressure from the AGN on dusty gas with $N_{\rm H} \geq 10^{22}$~cm$^{-2}$ is larger than the gravitational force \citep[e.g.,][]{Fabian2006,Fabian2008,Ishibashi2018}.}
As a result, they proposed a scenario that the covering fraction decreases with $\lambda_{\rm Edd}$ due to the radiation pressure \citep[see also][]{Ricci2022}.

It has been suggested that the AGNs in late mergers become heavily obscured by circumnuclear material.
The optical and IR observations have found that the AGNs in U/LIRGs show weak ionized lines from narrow line regions (NLRs), such as [\ion{O}{3}] $\lambda$5007
and [\ion{O}{4}] 25.89~$\mu$m \citep[e.g.,][]{Imanishi2006,Imanishi2008,Yamada2019}.
This type of AGNs is called ``buried'' AGNs, where no significant NLRs develop because the UV photons emitted from the accretion disk in almost all directions are absorbed by optically-thick ($A_{\rm V} \gtrsim 1$~mag or $N_{\rm H} \gtrsim 2 \times 10^{21}$~cm$^{-2}$; \citealt{Draine2003}) material.
Performing a broadband X-ray study with Nuclear Spectroscopic Telescope Array (NuSTAR) and Swift/BAT, \citet{Ricci2021b} found that \textcolor{black}{$f_{\rm obs}^{\rm (stat)}$($\geq$10$^{22}$~cm$^{-2}$) reaches $90\%$ for AGNs in local U/LIRGs.
The difference in fraction of Compton-thick (CT; with $N_{\rm H} \geq 10^{24}$~cm$^{-2}$) AGNs, $f_{\rm CT}^{\rm (stat)}$ = $f_{\rm obs}^{\rm (stat)}$(${\geq}10^{24}$~cm$^{-2}$), is more significant between late mergers ($\gtrsim$50\%) and Swift/BAT sources (20--30\%; \citealt{Ricci2015,Ricci2017cNature}).}
A multiwavelength study by \citet{Yamada2021} found that these AGNs in late mergers have high ${\log}\lambda_{\rm Edd}$ ($\gtrsim -$2).
How AGNs at high $\lambda_{\rm Edd}$, whose circumnuclear material should be affected by radiation pressure, get buried in dense obscurers remains an open question.
Most of the buried AGNs in U/LIRGs at $z \gtrsim 1$ would be missed in observations at any wavelengths. Hence, understanding the nature and evolution of ``buried'' AGNs may mean understanding a majority of the merger-induced AGNs in the Universe and thus the SMBH evolution in the context of cosmic history.

The first step of the study of buried AGNs is to identify the candidates and to construct its sample set, followed by a more detailed study, including to evaluate the relevant parameters and their relations, specifically the covering fraction and $\lambda_{\rm Edd}$.
\citet{Yamada2019} found a good indicator for identifying whether an AGN is buried or not, which employs the [\ion{O}{4}] 
line
to nuclear 12-$\mu$m luminosity ratio ($L_{\rm [O\,IV]}$/$L_{\rm 12,nuc}$).
A major downside in the method is the technical difficulty in measuring torus-originating $L_{\rm 12,nuc}$, which requires mid-IR observations with a high spatial resolution of subarcsecond to minimize the contamination from the host galaxy \citep{Asmus2014,Asmus2015}.
Accordingly, the available samples to obtain the indicator are limited to nearby, bright targets.

\textcolor{black}{Moreover, \citet{Ogawa2021} and \citet{Yamada2021} investigated the covering fractions based on the X-ray spectra, $f_{\rm obs}^{\rm (spec)}$(${\geq}10^{22}$~cm$^{-2}$), for the individual AGNs in Swift/BAT sources and U/LIRGs, respectively.
They performed X-ray spectroscopy with an X-ray torus model XCLUMPY \citep{Tanimoto2019}, which is one of the latest models based on the realistic geometries of clumpy tori \citep[e.g.,][]{Krolik1988,Wada2002,Honig2007,Laha2020}.
The XCLUMPY assumes Gaussian angular distributions of clumps \citep{Nenkova2008a,Nenkova2008b}, although we note that the $f_{\rm obs}^{\rm (spec)}$ depends on the assumption of an X-ray torus model.\footnote{\textcolor{black}{In XCLUMPY, the covering fraction of obscurers can be calculated by $f_{\rm obs}^{\rm (spec)}({\geq}N'_{\rm H}) = {\sin}(\sigma_{\rm X}({\ln(N_{\rm H}^{\rm Equ}/N'_{\rm H})})^{0.5}$), where $\sigma_{\rm X}$ is the torus angular width and $N_{\rm H}^{\rm Equ}$ is the equatorial hydrogen column density (see Figure~1 in \citealt{Tanimoto2019}; see also e.g., \citealt{Uematsu2021,Inaba2022,Nakatani2023}).
\citet{Tanimoto2018,Tanimoto2020,Tanimoto2022} and \citet{Buchner2019} report the difference in the X-ray spectral shape between the clumpy and smooth distribution of the obscurers.}}
For 28 Swift/BAT AGNs, \citet{Ogawa2021} estimated the values of $f_{\rm obs}^{\rm (spec)}$, which are well consistent with the values of $f_{\rm obs}^{\rm (stat)}$ at each $\lambda_{\rm Edd}$ bins reported by \citet{Ricci2017cNature}.
On the other hand, \citet{Yamada2021} found that the $f_{\rm obs}^{\rm (spec)}$($\geq$10$^{22}$~cm$^{-2}$) of AGNs with ${\log}\lambda_{\rm Edd} \gtrsim -$2 in late-merging U/LIRGs ($\sim$0.6--1.0) were significantly larger than the estimates for the Swift/BAT AGNs with ($f_{\rm obs}^{\rm (spec)} \sim 0.3$; \citealt{Ogawa2021}).
They discussed the covering fractions with $\geq$10$^{22}$~cm$^{-2}$, while in this study we focus on the covering fractions of CT obscurers, $f_{\rm CT}^{\rm (spec)}$ = $f_{\rm obs}^{\rm (spec)}$(${\geq}10^{24}$~cm$^{-2}$).
This is because the covering fractions of CT materials are less dependent on $\lambda_{\rm Edd}$ than those of absorbers with ${\geq}10^{22}$~cm$^{-2}$ among Swift/BAT AGNs;
accordingly, we can reveal a different factor that determines the covering fraction of CT material (e.g., merger stages).}

This work has two main goals. 
First, we establish useful diagnostics for identifying buried AGNs.
Specifically, we employ a similar method to our previous work \citep{Yamada2019} but using, together with [\ion{O}{4}], 12-$\mu$m AGN luminosities (designated as $L_{\rm 12,AGN}$;
e.g., \citealt{Yamada2023}) 
derived through spectral energy distribution (SED) analysis and also using [\ion{Ne}{5}] 14.32-$\mu$m data as a supplement.
The SED data are much more easily obtained than $L_{\rm 12,nuc}$.
To evaluate the $f^{\rm (spec)}_{\rm obs}$ for buried AGNs, we also compare the parameter with the $f^{\rm (stat)}_{\rm obs}$.
Second, we discuss the physical mechanism of the buried structure of AGNs in U/LIRGs by using covering fraction and $\lambda_{\rm Edd}$.
Section~\ref{S2-data} describes the sample we utilized.
In Section~\ref{S3-results}, we establish the diagnostics for buried AGNs.
Section~\ref{S4-discussion} discusses the structure of buried AGNs.
Section~\ref{S5-conclusion} summarizes the main results.
The notations of ${\log}N_{\rm H}$ and ${\log}L$ mean the logarithms of hydrogen column density in cm$^{-2}$ and luminosity in erg~s$^{-1}$, respectively.
All fractions are calculated by evaluating properly the binomial probability distribution with a beta function; we set the center value and its error range as the 50th and 16th--84th percentiles, respectively \citep{Cameron2011}.
\textcolor{black}{All errors of the parameters are 1$\sigma$ level and upper/lower limits are 3$\sigma$ level.}
Throughout the paper, we adopt a standard cosmology, 
$H_{\rm 0}$ = 70 km s$^{-1}$ Mpc$^{-1}$, $\Omega_{\rm M}$ = 0.3, and $\Omega_{\rm \Lambda}$ = 0.7.
\\


\section{Sample and Data Description}\label{S2-data}
For probing the buried AGNs with new diagnostics, we use two AGN samples in this work.
One is a sample set of the AGNs in \textcolor{black}{gas-rich mergers, selected from local U/LIRGs} \citep{Armus2009}.
The other is a sample of the \textcolor{black}{X-ray-selected AGNs in nonmerging galaxies, detected with Swift/BAT} \citep{Ricci2017dApJS}.
The AGNs contained in the two samples are 
\textcolor{black}{exclusive}.

\begin{figure*}
    \centering
    \includegraphics[keepaspectratio, scale=0.52]{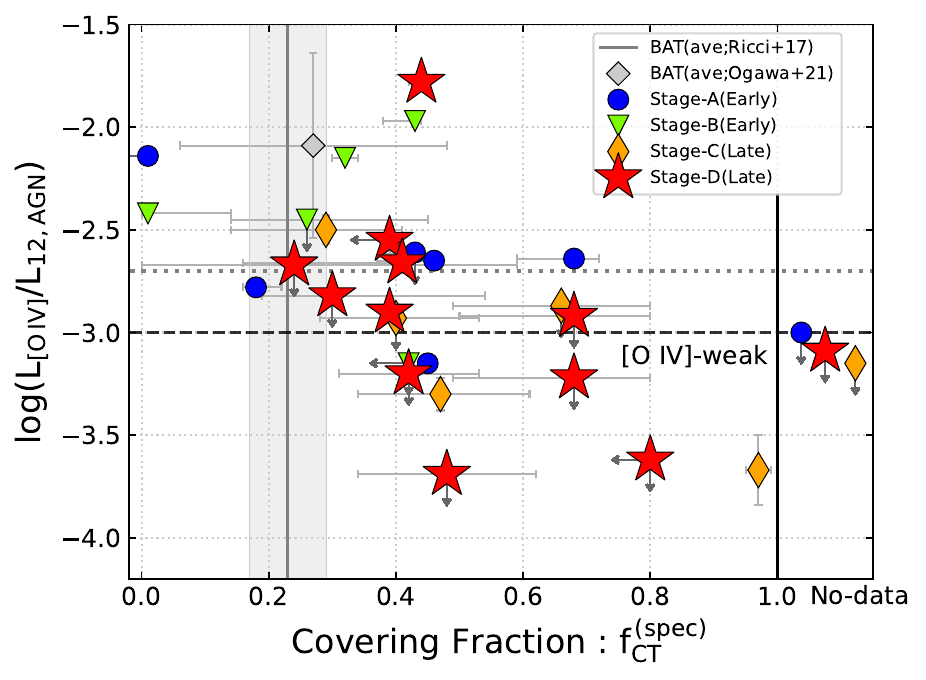}
    \includegraphics[keepaspectratio, scale=0.52]{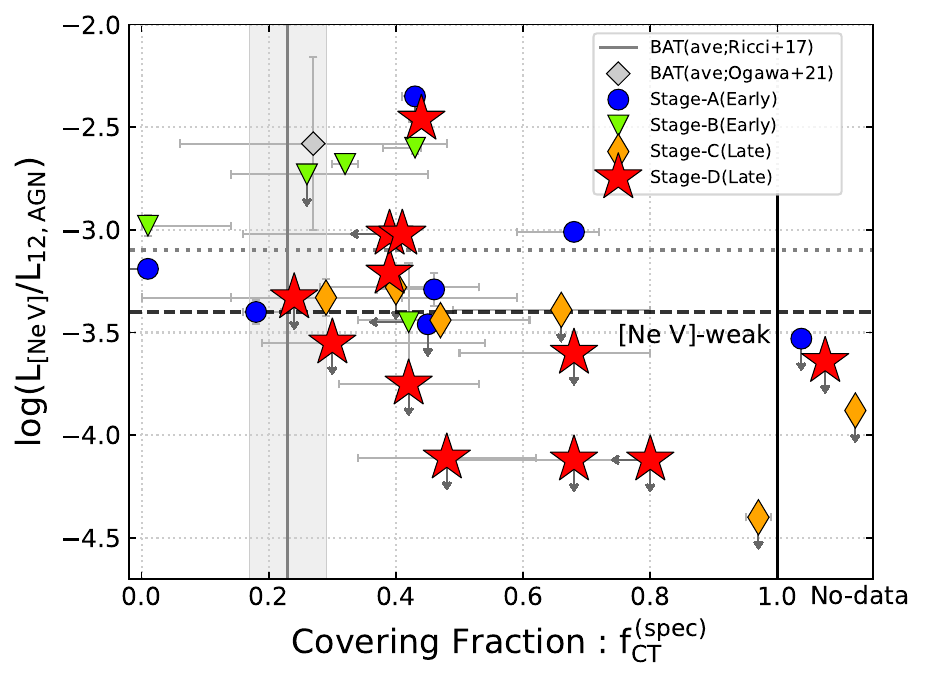}
    \caption{Left panel: ${\log}(L_{\rm [O\,IV]}$/$L_{\rm 12,AGN})$ vs. covering fraction of CT material ($f^{\rm (spec)}_{\rm CT}$).
    Horizontal dashed and dotted lines mark the thresholds of [\ion{O}{4}]-weak AGNs for ${\log}L_{\rm [O\,IV]}$/$L_{\rm 12,AGN} \leq -$3.0 and $-$2.7 (the latter is for data points with only upper limits given), respectively.
    The vertical solid line and gray shaded area show the typical value with a 1$\sigma$ dispersion for the Swift/BAT AGNs from \citet{Ricci2017cNature}, assuming $f^{\rm (spec)}_{\rm CT}$ = $f^{\rm (stat)}_{\rm CT}$.
    The gray diamond is the average of the 14 BAT-IRS AGNs in \citet{Ogawa2021}.
    The other symbols denote the merger stages as described in the legend at the top-right corner.
    \textcolor{black}{Right panel: the same as the left but for [\ion{Ne}{5}] line.
    Horizontal dashed and dotted lines are the thresholds of [\ion{Ne}{5}]-weak AGNs for ${\log}L_{\rm [Ne\,V]}$/$L_{\rm 12,AGN} \leq -$3.4 and $-$3.1 (the latter is for upper limits), respectively.}
    \\}
\label{F1-oiv-nev-12um}
\end{figure*}

\subsection{GOALS Sample}\label{sub2-1-GOALS}
Our base sample of U/LIRGs is the Great Observatories All-Sky LIRG Survey (GOALS; \citealt{Armus2009}).
It consists of 202 U/LIRGs at $z < 0.088$, which are selected from flux-limited IR sources detected with IRAS.
\textcolor{black}{Among these objects, \citet{Yamada2021} identified 57 U/LIRG systems that had been observed with hard X-rays, which totaled 84 individual nuclei in the optical band and found significant hard X-ray detection from 40 AGNs from these nuclei.
\citet{Yamada2023} performed hard-X-ray-to-radio SED decomposition for these 57 U/LIRGs (or 72 component sources resolved in the Herschel 70~$\mu$m band).
They cataloged the intrinsic AGN luminosities for the 40 hard X-ray-detected AGNs in the multiwavelength bands, containing two Herschel-unresolved dual-AGN systems and 36 resolved single AGNs.} 
The study further identified three AGN candidates through a significant test for two kinds of SED fits (with and without an AGN component)
with a reduced Bayesian information threshold of equal to or more than 6.0 (i.e., posterior probability above 95\%; \citealt{Raftery1995}).
The three AGN candidates should be CT AGNs on the basis of their 12-$\mu$m and X-ray luminosities (see Appendix~\ref{AppendixA1-CTAGN}).
\textcolor{black}{Here, we exclude two unresolved dual-AGN systems (Mrk~266 and NGC~6240), two AGNs whose host galaxies have much smaller $L_{\rm IR}$ ($<$10$^{11} L_{\odot}$; NGC 6921 and NGC 7682) than those of the interacting pairs, and seven nonmerging sources (see Appendix~\ref{AppendixA2-nonmerger}). Finally, our sample set consists of 30 sources, consisting of 27 hard X-ray-detected AGNs and three CT AGN candidates.}

\citet{Yamada2021} estimated the covering fractions of their cataloged sources with XCLUMPY and presented their SMBH masses and merger stages in four levels of A, B (early mergers), C, and D (late mergers), classified on the basis of their high-spatial-resolution optical/infrared images.
The [\ion{O}{4}] and [\ion{Ne}{5}] fluxes were measured with Spitzer/IRS (\citealt{Inami2013}; except for IC 4518, which is taken from \citealt{Spoon2022}).
\textcolor{black}{In Table~\ref{T1-AGN-property}, we summarize the main properties of the 27 AGNs and three AGN candidates in merging U/LIRGs.}

\subsection{Swift/BAT AGN Sample}\label{sub2-2-BAT-AGN}
We construct our sample set of X-ray-selected, Swift/BAT-Spitzer/IRS non-blazer AGNs from the 606 AGNs selected by \citet{Ichikawa2017} from the Swift/BAT 70-month catalog. The combined selection criteria by \citet{Ichikawa2017} and us are:
(i) galactic latitudes ($|b| > 10$\degr),
(ii) spectroscopic redshifts being available,
\textcolor{black}{(iii) not interacting galaxies \citep{Koss2012,Yamada2021},
(iv) both [\ion{O}{4}] and [\ion{Ne}{5}] line luminosities being constrained with Spitzer/IRS,
(v) 12-$\mu$m AGN luminosity being constrained,
and (vi) SMBH mass and Eddington ratios being presented in \citet{Koss2022b}.
The resultant BAT-IRS sample contains 138 AGNs in nonmergers.} 
Appendix~\ref{AppendixB-BAT} describes the detailed selection process.

So far, \textcolor{black}{27 of the 138 AGNs have been studied by our group using the XCLUMPY model.
Thirteen AGNs are in \citet{Tanimoto2022},} who systematically analyzed CT AGN candidates that may have dense circumnuclear material (see Section~\ref{sub4-1-origin}).
Except for these AGNs, the covering fractions of the rest of the 14 non-CT AGNs studied in \citet{Ogawa2021}
are available and are consistent with the relation between $\lambda_{\rm Edd}$ and covering fraction derived from the obscured AGN fraction \citep{Ricci2017cNature}.
Although the 14 AGNs are closer objects ($z < 0.02$) than the 138 AGNs ($z \sim 0.04$ on average), they cover a wide range of ${\log}\lambda_{\rm Edd}$ ($-$3--0).
A Kolmogorov-Smirnov (KS) test indicates that the difference in ${\log}\lambda_{\rm Edd}$ between BAT-IRS AGNs and 14 non-CT AGNs is not significant ($p$-value = 0.72).
Thus, we treat the 14 non-CT AGNs studied with XCLUMPY as representative ones (see also Section~\ref{sub3-1-oiv-weak}). Table~\ref{T1-AGN-property} lists their properties.
\\


\section{Results}\label{S3-results}
We propose a new diagnostic method to determine to what degree individual AGNs are ``buried'', i.e., in what covering fraction ($f^{\rm (spec)}_{\rm CT}$) CT matter covers the central AGN, by using the ratio of the [\ion{O}{4}] luminosity to 12-$\mu$m AGN luminosity derived on the basis of IR SED analysis ($L_{\rm [O\,IV]}$/$L_{\rm 12,AGN}$; see Section~\ref{S1-intro} and \citealt{Yamada2023}, for detail). 
We define the [\ion{O}{4}]-weak AGNs and report their fraction in the AGNs in the GOALS and BAT-IRS samples.
\textcolor{black}{The ratio of the [\ion{Ne}{5}] luminosity ($L_{\rm [Ne\,V]}$) to $L_{\rm 12,AGN}$ is also used as a tracer of buried AGNs.}

\begin{figure*}
    \centering
    \includegraphics[keepaspectratio, scale=0.42]{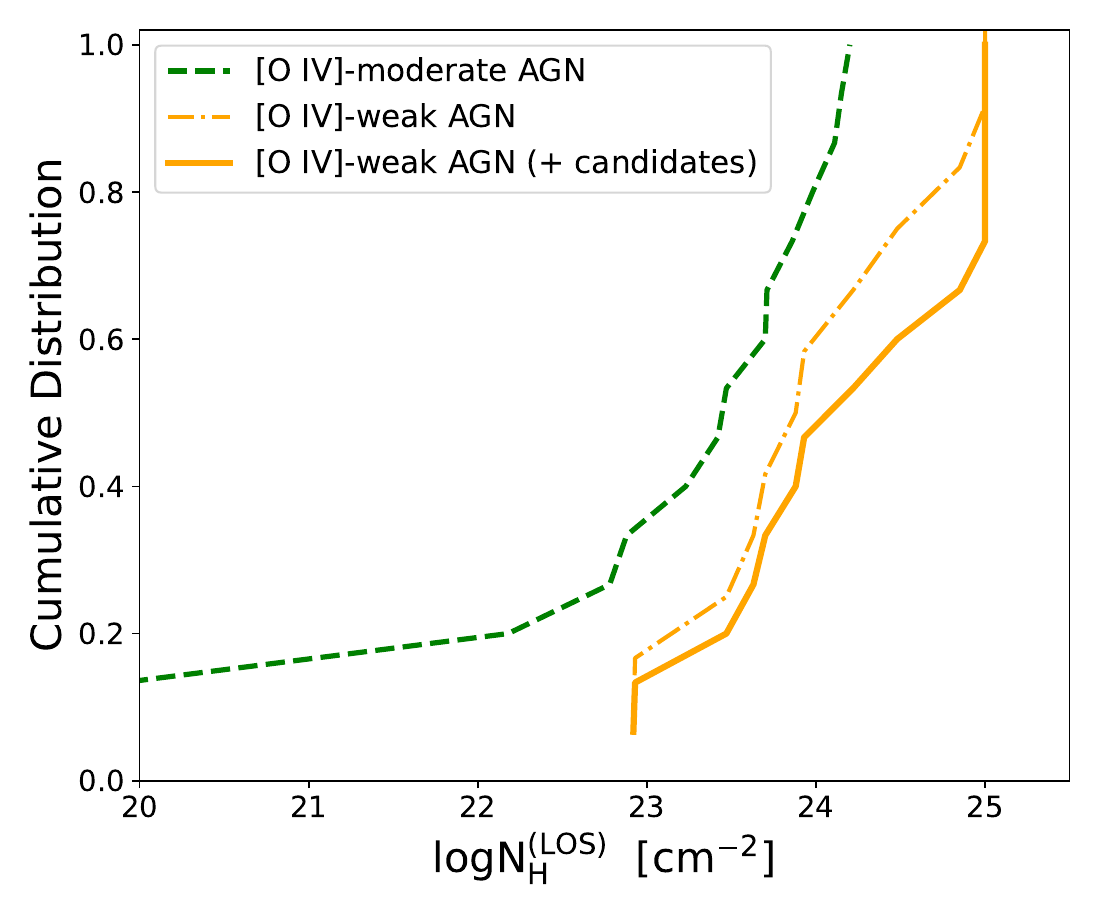}
    \includegraphics[keepaspectratio, scale=0.42]{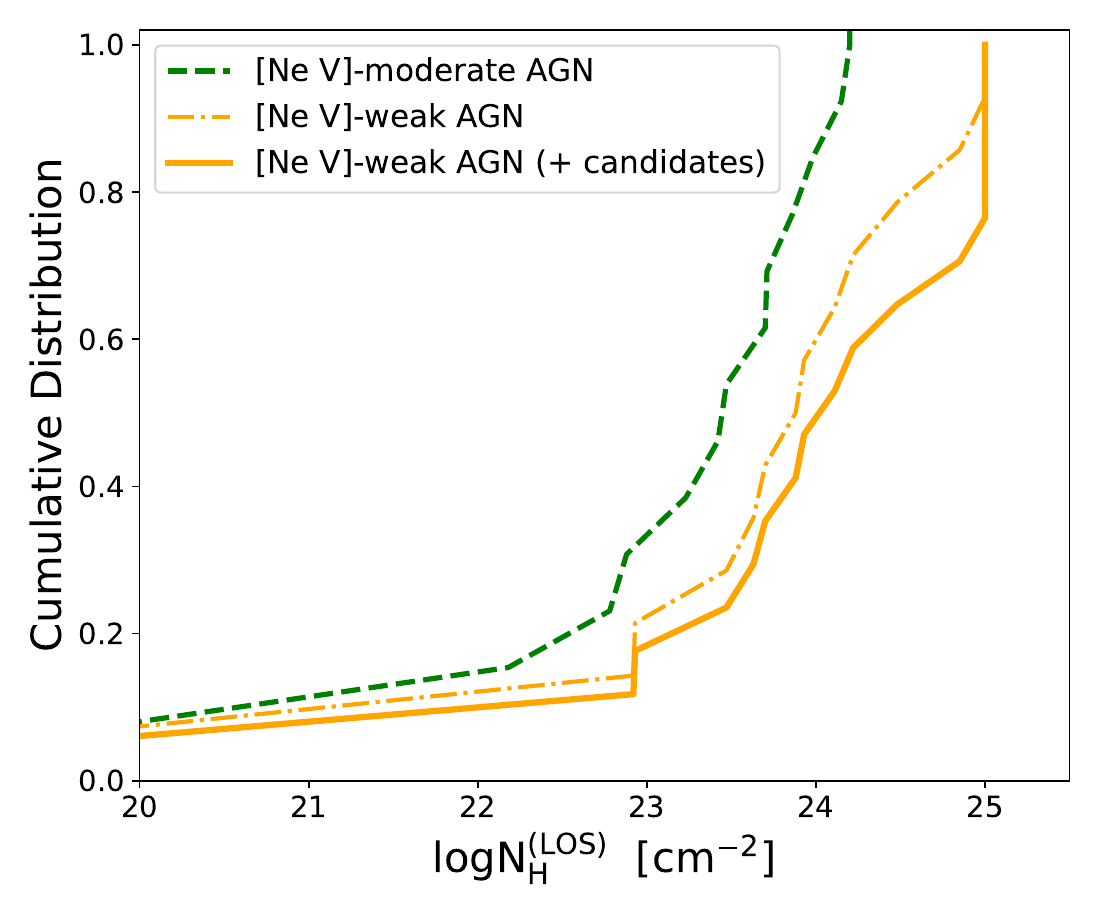}
    \caption{Left panel: cumulative $N_{\rm H}$ distribution for (green dashed line) [\ion{O}{4}]-moderate AGNs, (orange dashed line) [\ion{O}{4}]-weak AGNs, and (orange solid line) [\ion{O}{4}]-weak AGNs plus three CT AGN candidates among merging U/LIRGs.
    \textcolor{black}{Right panel: the same as the left but for [\ion{Ne}{5}] line.}
    \\}
\label{F2-cum-nh}
\end{figure*}

\subsection{\textcolor{black}{Definition of [\ion{O}{4}]-weak and [\ion{Ne}{5}]-weak AGNs}}\label{sub3-1-oiv-weak}
We make a scatter plot of $L_{\rm [O\,IV]}$/$L_{\rm 12,AGN}$ and $f^{\rm (spec)}_{\rm CT}$ estimated for the X-ray data of our sample with XCLUMPY (left panel of Figure~\ref{F1-oiv-nev-12um}) to search for a potential correlation with the aim of using it as a tracer to evaluate the nature of potentially buried AGNs.
We then perform regression analysis of the data, using the Bayesian maximum-likelihood method (\citealt{Kelly2007}; see also \citealt{Toba2019a}), which can handle even poorly constrained data with only upper limits determined, and find that the AGNs in U/LIRGs show a significant anti-correlation for the two quantities with a correlation coefficient of $-0.48 \pm 0.19$.
Its $p$-value is 0.0019, indicating that the correlation is significant at a ${>}99$\% confidence level.
This result supports that $L_{\rm [O\,IV]}$/$L_{\rm 12,AGN}$ can be a useful tracer of $f^{\rm (spec)}_{\rm CT}$.
Also, we find that the average values of $L_{\rm [O\,IV]}$/$L_{\rm 12,AGN}$ and $f^{\rm (spec)}_{\rm CT}$ of the 14 BAT-IRS AGNs (Section~\ref{sub2-2-BAT-AGN}; gray diamond), $-2.09 \pm 0.45$ and $0.27 \pm 0.21$, respectively, are within the typical range of the AGNs in U/LIRGs.

\textcolor{black}{Similarly, we investigate a correlation between $L_{\rm [Ne\,V]}$/$L_{\rm 12,AGN}$ and $f^{\rm (spec)}_{\rm CT}$ (right panel of Figure~\ref{F1-oiv-nev-12um}).
The correlation coefficient is $-0.36 \pm 0.24$ and its $p$-value is 0.020 ((i.e., a ${>}95$\% confidence level).
The average values of $L_{\rm [Ne\,V]}$/$L_{\rm 12,AGN}$ of the 14 BAT-IRS AGNs in \citet{Ogawa2021} is $-2.58 \pm 0.42$.
Although the significance of the correlation is not strong, the $L_{\rm [Ne\,V]}$/$L_{\rm 12,AGN}$ can also be an indicator of $f^{\rm (spec)}_{\rm CT}$.}

\textcolor{black}{To define criteria of [\ion{O}{4}]-weak and [\ion{Ne}{5}]-weak AGNs, from the 138 BAT-IRS AGNs we here focus on 86 AGNs both of whose [\ion{O}{4}] and [\ion{Ne}{5}] fluxes and the 1$\sigma$ errors are measured.
For the other BAT-IRS AGNs, the 3$\sigma$ upper limits of the line fluxes are constrained.
In the $L_{\rm [O\,IV]}$/$L_{\rm 12,AGN}$ and $L_{\rm [Ne\,V]}$/$L_{\rm 12,AGN}$ distributions for the 86 AGNs, the lower side of a 90\% confidence interval corresponds to $L_{\rm [O\,IV]}$/$L_{\rm 12,AGN} \lesssim -$3.0 and $L_{\rm [Ne\,V]}$/$L_{\rm 12,AGN} \lesssim -$3.4, respectively.
This threshold, however, should be an overestimate for the data points (i.e., AGNs) with only upper limits determined, containing the other BAT-IRS AGNs.}
Since the typical 1$\sigma$ error is $\sim$0.1~dex, we consider 0.3~dex as systematic errors,
which is translated into the threshold of ${\log}L_{\rm [O\,IV]}^{\rm (upper)}$/$L_{\rm 12,AGN} \lesssim -2.7$, for those poorly determined data points.
In consequence, we define AGNs with ${\log}L_{\rm [O\,IV]}$/$L_{\rm 12,AGN} \leq -3.0$ (or $-2.7$ if only upper limits are given for $L_{\rm [O\,IV]}$) as [\ion{O}{4}]-weak and the others as [\ion{O}{4}]-moderate.
\textcolor{black}{Similarly, we define [\ion{Ne}{5}]-weak AGNs by ${\log}L_{\rm [Ne\,V]}$/$L_{\rm 12,AGN} \leq -3.4$ (or $-3.1$ for upper limits).
Appendix~\ref{AppendixC-uplim} discusses the case of universal adoption of the threshold of ${\log}L_{\rm [O\,IV]}$/$L_{\rm 12,AGN} \leq -3.0$ and ${\log}L_{\rm [Ne\,V]}$/$L_{\rm 12,AGN} \leq -3.4$ for upper limits, which is technically more straightforward and is a viable option if the systematic uncertainty is small.}

\subsection{Cumulative $N_{\rm H}$ Distribution for Buried AGNs}\label{sub3-2-ULIRGs}
Under the definition of the [\ion{O}{4}] weakness in the previous subsection, 12 out of 27 hard X-ray-detected AGNs in U/LIRGs are classified into [\ion{O}{4}]-weak AGNs. 
For the [\ion{O}{4}]-weak ones, the median and average of $f^{\rm (spec)}_{\rm CT}$ estimated with XCLUMPY are $0.48\pm0.13$ and $0.55\pm0.19$, respectively.
These are significantly larger than the fraction of CT AGNs for Swift/BAT AGNs ($f^{\rm (stat)}_{\rm CT} = 23 \pm 6$\%; \citealt{Ricci2017cNature}).
This implies that a high fraction of [\ion{O}{4}]-weak AGNs are buried AGNs.

\textcolor{black}{To assess the covering fractions estimated with XCLUMPY in an independent way, we compare the cumulative $N_{\rm H}$ distributions for [\ion{O}{4}]-weak and [\ion{O}{4}]-moderate AGNs in merging U/LIRGs in the left panel of Figure~\ref{F2-cum-nh}.
We find that [\ion{O}{4}]-weak AGNs for sources without three AGN candidates (orange dashed-dotted line in Figure~\ref{F2-cum-nh}) have significantly larger $N_{\rm H}$ than [\ion{O}{4}]-moderate ones as a whole (green dashed line).
Moreover, among the 15 [\ion{O}{4}]-weak sources, containing three plausible CT AGN candidates (see Appendix~\ref{AppendixA1-CTAGN}), which are [\ion{O}{4}]-weak, the $N_{\rm H}$ distribution becomes larger (solid orange line).
Eight of them are CT AGNs, corresponding to a fraction of $53\pm12$\% (8/15, i.e., 8 out of 15 AGNs).}
The fraction is consistent with the above-mentioned XCLUMPY-estimated average ($\sim$0.55), but is much larger than 
the CT AGN fraction for the [\ion{O}{4}]-moderate AGNs ($29^{+11}_{-10}$\%; 4/15, see Table~\ref{T1-AGN-property}). 
\textcolor{black}{For the [\ion{O}{4}]-weak AGNs, the fraction of AGNs with $N_{\rm H} \geq 10^{22}$~cm$^{-2}$ is $f^{\rm (stat)}_{\rm obs}$(${\geq}10^{22}~{\rm cm}^{-2}) = 96^{+3}_{-7}$\% (15/15).
Similarly, we find 17 [\ion{Ne}{5}]-weak AGNs, containing all the 15 [\ion{O}{4}]-weak AGNs and two [\ion{O}{4}]-moderate AGNs with ${\log}L_{\rm [O\,IV]}$/$L_{\rm 12,AGN} \leq -2.67$ (NGC~7469 and IRAS~F14348$-$1447).
They show a similar $N_{\rm H}$ distribution in the right panel of Figure~\ref{F2-cum-nh}.
The covering factors of the [\ion{Ne}{5}]-weak AGNs are $f^{\rm (stat)}_{\rm CT} = 53^{+11}_{-12}$\% (9/17) and $f^{\rm (stat)}_{\rm obs}({\geq}10^{22}~{\rm cm}^{-2}) = 91^{+5}_{-8}$\% (16/17).
Although the number of the [\ion{Ne}{5}]-filtered buried AGNs might be somewhat larger than that of the [\ion{O}{4}]-filtered, these imply that the population of [\ion{O}{4}] and [\ion{Ne}{5}]-weak AGNs have large CT covering fractions and are actually ``buried'' by optically-thick (${\gtrsim}2 \times 10^{21}$~cm$^{-2}$; see Section~\ref{S1-intro}) material in almost all directions.}

\begin{figure}
    \centering
    \includegraphics[keepaspectratio, scale=0.53]{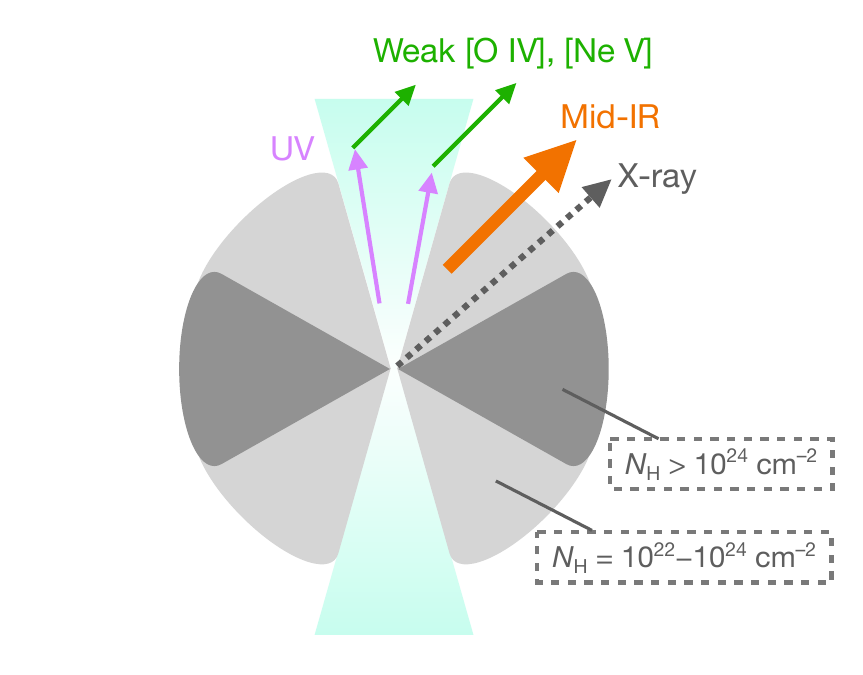}
    \caption{\textcolor{black}{Schematic picture of the structure of buried AGNs.
    The covering fractions of obscurers are very high, $f^{\rm (spec)}_{\rm obs} ({\geq}10^{22}$~cm$^{-2}) \gtrsim 0.9$ and $f^{\rm (spec)}_{\rm CT} \sim 0.5$.
    The ionized line emission from NLRs (e.g., [\ion{O}{4}] and [\ion{Ne}{5}]) becomes weak because the UV photons are absorbed in almost all directions by optically thick material.}
    \\}
\label{F3-buried-AGN}
\end{figure}

The derived fractions $f^{\rm (stat)}_{\rm CT}$ ($\sim f^{\rm (spec)}_{\rm CT}$) and cumulative distribution of [\ion{O}{4}] and [\ion{Ne}{5}]-weak AGNs in U/LIRGs are similar to those of the AGNs in late mergers that are thought to be buried \citep[e.g.,][]{Ricci2017bMNRAS,Ricci2021b,Yamada2021}.
To conclude, the threshold of [\ion{O}{4}] and [\ion{Ne}{5}] weakness enables us to select buried AGNs with $f^{\rm (spec)}_{\rm CT} \sim 0.5 \pm 0.1$, which may intrinsically differ from the Swift/BAT AGNs.
\textcolor{black}{Based on the $N_{\rm H}$ cumulative distribution and an equation of $f_{\rm obs}^{\rm (stat)}$ = $(1/4\pi) \int_{0}^{2\pi} \int_{-\theta}^{\theta} sin(\theta')d\theta' d\phi $ = cos$\theta$ ($\theta$: the half opening angle), the structure of [\ion{O}{4}]-weak or [\ion{Ne}{5}]-weak AGNs can be described in Figure~\ref{F3-buried-AGN}.}

\begin{figure*}
    \centering
    \includegraphics[keepaspectratio, scale=0.4]{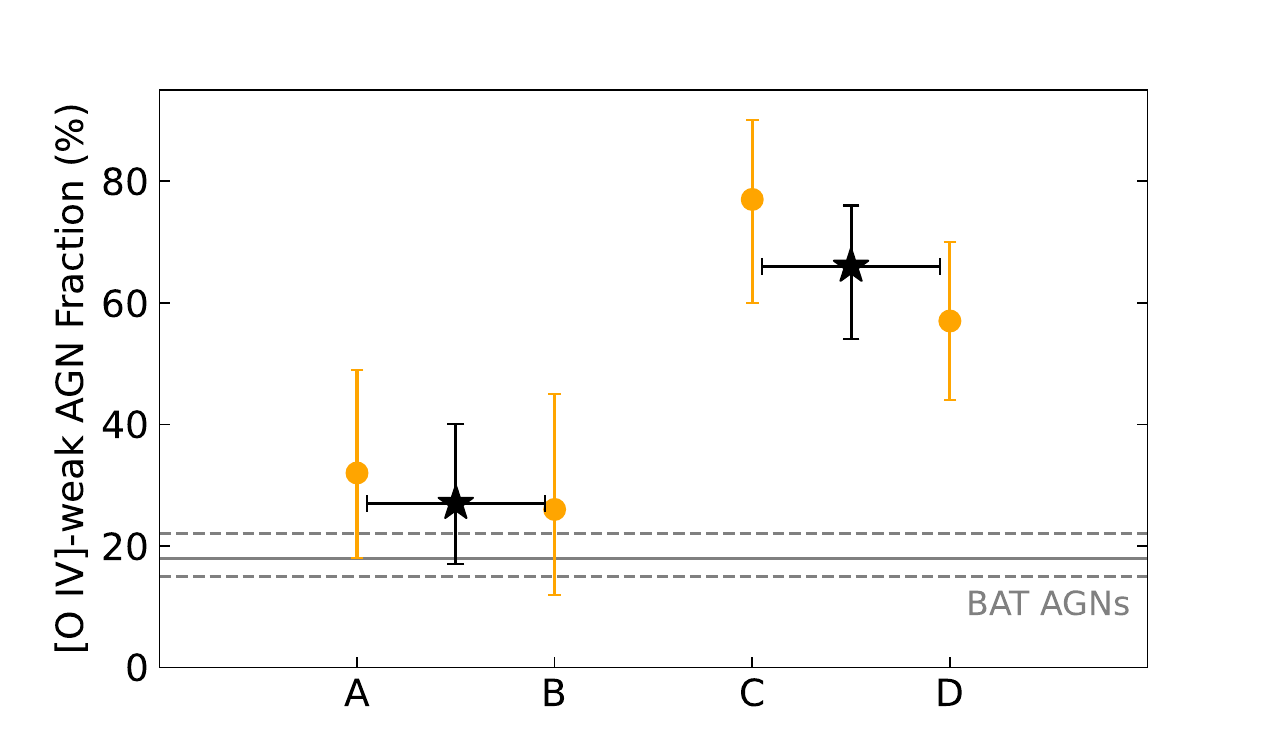}
    \includegraphics[keepaspectratio, scale=0.4]{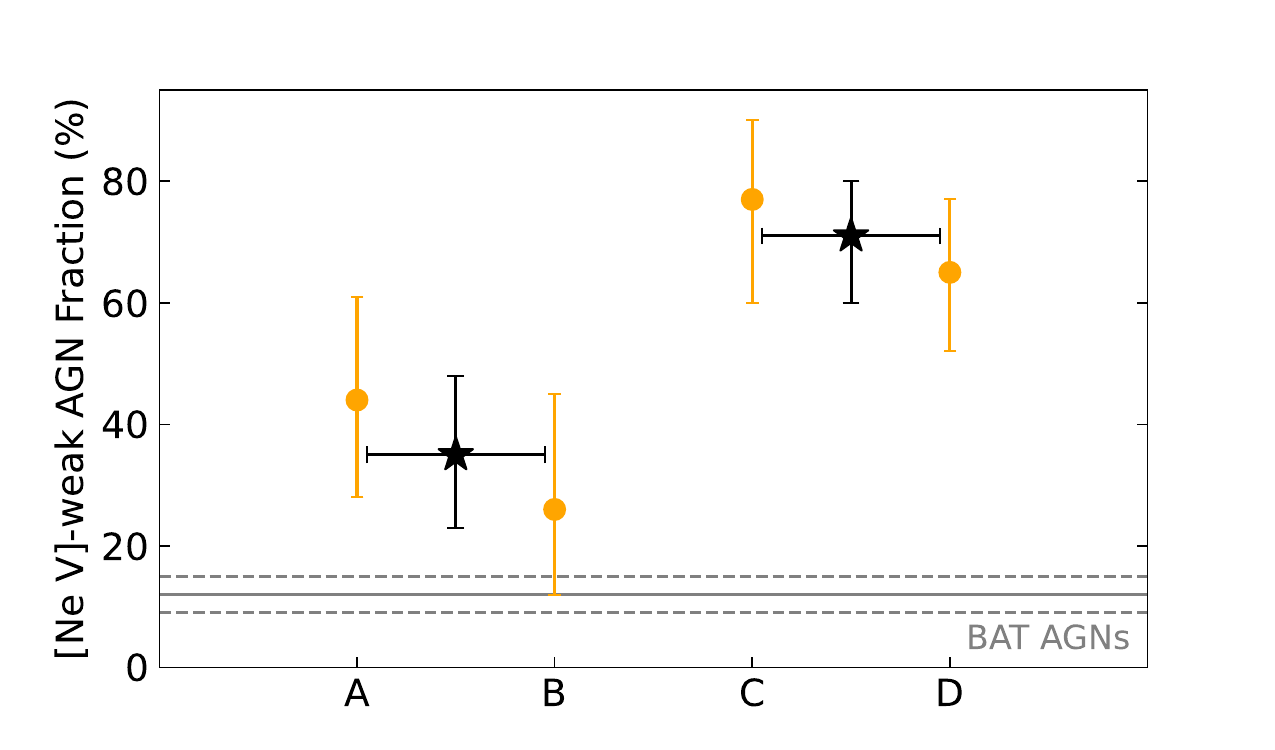}
    \caption{Left panel: fraction of [\ion{O}{4}]-weak AGNs by merger stage.
    Orange circles indicate those for stages A--D, and black stars, for early and late mergers.
    The gray solid and dashed lines mark the average with 1$\sigma$ dispersion for 138 BAT-IRS AGNs.
    \textcolor{black}{Right panel: the same as the left but for [\ion{Ne}{5}] line.}
    \\}
\label{F4-buried-fraction}
\end{figure*}

\subsection{Fraction of Buried AGNs}\label{sub3-3-oivweak-fraction}
\textcolor{black}{Since $f^{\rm (spec)}_{\rm CT}$ depends on the assumption of the X-ray torus model (e.g., clumpy or smooth distribution)}, we examine the fraction of [\ion{O}{4}]-weak AGNs according to merger stages in U/LIRGs (left panel of Figure~\ref{F4-buried-fraction}) and test the consistency between $f^{\rm (spec)}_{\rm CT}$ and $f^{\rm (stat)}_{\rm CT}$.
The fractions of [\ion{O}{4}]-weak AGNs are $27^{+13}_{-10}$\% (3/12) and $66^{+10}_{-12}$\% (12/18) in early and late mergers, respectively.
Similar results can be seen for [\ion{Ne}{5}]-weak AGNs. 
The fractions of [\ion{Ne}{5}]-weak AGN are $35^{+13}_{-12}$\% (4/12; early) and $71^{+9}_{-11}$\% (13/18; late mergers), as shown in the right panel of Figure~\ref{F4-buried-fraction}.
These fractions in early and late mergers are larger than $18^{+4}_{-3}$\% (25/138 for [\ion{O}{4}] line) or $12 \pm 3$\% (16/138 for [\ion{Ne}{5}] line) in the BAT-IRS AGNs.
Adopting $f^{\rm (spec)}_{\rm CT} = 0.4$--0.6 (for [\ion{O}{4}]-weak) and 0.2--0.3 (for the others), the expected fraction of CT AGNs 
(= sum of the fraction of [\ion{O}{4}]-weak/moderate AGNs times each $f^{\rm (spec)}_{\rm CT}$)
should be $f^{\rm (stat)}_{\rm CT} = 20$--40\% and 30--50\% for early and late mergers, respectively.
The values agree with the actual detection rates of CT AGNs at the respective merger stages, $f^{\rm (stat)}_{\rm CT} \sim 30$\% and $\sim$50\% \citep[e.g.,][]{Ricci2021b,Yamada2021}.
\textcolor{black}{The independent results of $f^{\rm (spec)}_{\rm CT}$ and $f^{\rm (stat)}_{\rm CT}$ are well consistent.
We also find that} the fraction of the buried AGNs in U/LIRGs increases with the merger stage from $\sim$27\% (early) to $\sim$66\% (late mergers).

\textcolor{black}{The fraction of obscured AGNs among [\ion{O}{4}]-weak (or [\ion{Ne}{5}]-weak) AGNs is very different between the merging U/LIRGs and BAT-IRS sources.
The fraction of obscured AGNs among [\ion{O}{4}]-weak BAT-IRS AGNs is $f^{\rm (stat)}_{\rm obs}$(${\geq}10^{22}$~cm$^{-2}) = 29^{+9}_{-8}$\% (7/25), much smaller than that of [\ion{O}{4}]-weak ones in U/LIRGs ($96^{+3}_{-7}$\%; Section~\ref{sub3-2-ULIRGs}).
The value for [\ion{Ne}{5}] line is also small ($33 \pm 11$\%; 5/16).
A possible hypothesis for the origin of less obscured sources is that NLRs may have not been developed due to too a short time after a recent trigger of the AGNs.
These AGNs may be contained in U/LIRG sample, but the larger fraction of [\ion{O}{4}] or [\ion{Ne}{5}]-weak AGNs (Figure~\ref{F4-buried-fraction}) and higher $f^{\rm (stat)}_{\rm obs}$(${\geq}10^{22}$~cm$^{-2}$) in U/LIRGs suggest that the [\ion{O}{4}] or [\ion{Ne}{5}]-weak AGNs are likely different from those in the BAT-IRS sources.
To unveil the origin of [\ion{O}{4}] or [\ion{Ne}{5}]-weak BAT-IRS AGNs, many of which are not likely buried, further studies are needed to draw a conclusion.}

By contrast, \textcolor{black}{the 10 obscured [\ion{O}{4}] or [\ion{Ne}{5}]-weak BAT-IRS AGNs with $N_{\rm H} \geq 10^{22}$~cm$^{-2}$ can be buried AGN candidates.
Their properties are listed in Table~\ref{T1-AGN-property}.
Six AGNs out of them have $N_{\rm H} \geq 10^{23}$~cm$^{-2}$ and their fractions of scattered light from the AGN with respect to the intrinsic AGN component are extremely small ($f_{\rm scat} < 0.5$\%; \citealt{Ricci2017dApJS}; see also \citealt{Ueda2007,Kawamuro2016}).
The estimates of $f_{\rm scat}$ are 0.2--0.3\% (NGC~4992, ESO 297$-$18, Z~147$-$20, ESO 506$-$27), ${\leq}1.3$\% (NGC~7479), and ${\leq}5.2$\% (Mrk~622), while $f_{\rm scat} > 0.5$\% for the other four less-obscured AGNs.}
The low-X-ray-scattering AGNs are difficult to find for U/LIRGs due to the
contamination from hot gas and X-ray binaries in the starburst regions, while they were found in Swift/BAT AGNs \citep[e.g.,][]{Yamada2019}.
The obscured AGNs with low $f_{\rm scat}$ in soft X-rays are known to show on average smaller [\ion{O}{3}] line to X-ray luminosity ratios ($L_{\rm [O\,III]}$/$L_{\rm X}$) than the other (i.e., $f_{\rm scat} \ge 0.5$\%) obscured AGNs \citep{Ueda2015,Gupta2021}, the fact of which suggests that the former population of AGNs is buried in circumnuclear material with small opening angles. 
Therefore, the six AGNs are likely to be buried AGNs.
The fraction of [\ion{O}{4}] or [\ion{Ne}{5}]-weak AGNs with $N_{\rm H} \geq 10^{23}$~cm$^{-2}$ is only $5 \pm 2$\% (6/138) for the BAT-IRS AGNs.
Considering that the fraction is much smaller than that of buried AGNs in late mergers ($66^{+10}_{-12}$\%), galaxy merging is likely to enhance the buried structure. 
\textcolor{black}{We have finally found 23 buried AGN candidates, containing 17 [\ion{O}{4}] or [\ion{Ne}{5}]-weak AGNs in merging U/LIRGs (where 4 are early mergers and 13 are late mergers) and 6 obscured ones in the nonmerging BAT-IRS AGNs.}
\\


\section{Discussion}\label{S4-discussion}

\subsection{How Do AGNs Get Buried in U/LIRGs?}\label{sub4-1-origin}
For unveiling the origin of the buried structure in late mergers, we examine the effect of radiation pressure on the obscurers by focusing on the effective Eddington ratio ($\lambda_{\rm Edd}^{\rm eff}$).
\textcolor{black}{The ratio $\lambda_{\rm Edd}^{\rm eff}$, calculated with the radiation pressure against the gravitational force by the SMBH, is an Eddington limit for dusty gas, in which the scattering cross-section for UV photons is taken into account \citep[e.g.,][]{Fabian2008}.
\citet{Ishibashi2018} calculate the $\lambda_{\rm Edd}^{\rm eff}$, taking account of the effect of not only UV absorption by dusty gas but also their IR re-radiation.}
Recent studies of Swift/BAT AGNs \citep{Ricci2022,Ananna2022} found that the $f_{\rm obs}^{\rm (stat)}$($\geq$10$^{22}$~cm$^{-2}$) decreases steeply at ${\log}\lambda_{\rm Edd} \gtrsim -$2, corresponding to $\lambda_{\rm Edd}^{\rm eff}$ for the obscurers with $N_{\rm H} \geq 10^{22}$~cm$^{-2}$.
This supports that their circumnuclear material is blown away by the radiation pressure when $\lambda_{\rm Edd} > \lambda_{\rm Edd}^{\rm eff}$.
\citet{Ricci2017cNature} report that the covering fractions of CT material in the BAT AGNs hardly vary, $f^{\rm (stat)}_{\rm CT} \sim 23$\%, which they argue may be mainly because the CT material at a location of a small elevation angle from the disk suffer a much smaller radiation pressure than that in regions closer to the polar angle and thus are not easily blown away and remain there, \textcolor{black}{constituting the persistent CT region (e.g., \citealt{Wada2015}).}

In contrast, \citet{Yamada2021} found in their U/LIRG sample set that some AGNs in late mergers with ${\log}\lambda_{\rm Edd} \gtrsim -$2 
have larger $f_{\rm obs}^{\rm (stat)}$ and $f_{\rm obs}^{\rm (spec)}$ ($\geq$10$^{22}$~cm$^{-2}$) than those of Swift/BAT AGNs with the similar $\lambda_{\rm Edd}$ \citep{Ricci2017cNature,Ogawa2021}.
Thus, these studies indicate that the AGNs in late mergers 
are not likely to follow the same trend regulated by the AGN-driven radiation pressure for obscurers with $N_{\rm H} \geq 10^{22}$~cm$^{-2}$ as for Swift/BAT AGNs.

\subsubsection{Relation between the CT covering fraction and $\lambda_{\rm Edd}$}\label{subsub4-1-1-CT-Edd}
\begin{figure}
    \centering
    \includegraphics[keepaspectratio, scale=0.54]{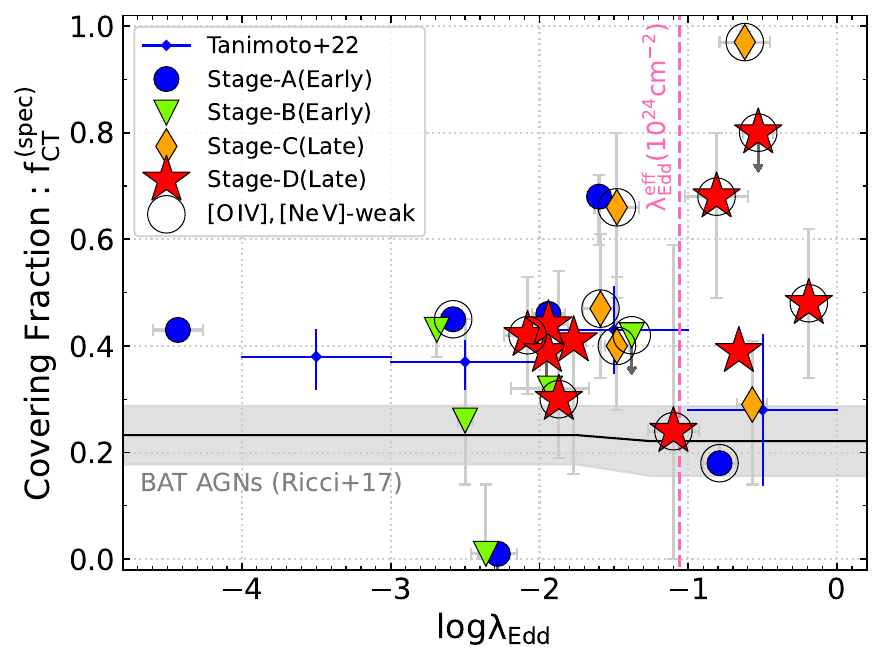}
    \caption{$f^{\rm (spec)}_{\rm CT}$ derived from X-ray studies with XCLUMPY vs. $\lambda_{\rm Edd}$ for merging U/LIRGs.
    The black solid curve and gray shaded area indicate the typical relation with a 1$\sigma$ dispersion for Swift/BAT AGNs \citep{Ricci2017cNature}, assuming $f^{\rm (spec)}_{\rm CT}$ = $f^{\rm (stat)}_{\rm CT}$.
    The vertical pink dashed line marks the $\lambda_{\rm Edd}^{\rm eff}$ for CT material \citep{Ishibashi2018}. 
    Small blue diamonds with uncertainties show the mean values and standard errors in the four log$\lambda_{\rm Edd}$ bins for the 52 CT AGN candidates, taken from \citet[][see text for detail]{Tanimoto2022}.
    The data points of the [\ion{O}{4}]-weak or [\ion{Ne}{5}]-weak AGNs are circled in black.
    The notations for the other symbols are the same as in Figure~\ref{F1-oiv-nev-12um}.
    \\}
\label{F5-CT-Edd}
\end{figure}

To understand the origin of the heavily obscured structure in buried AGNs, we examine the relation between $f^{\rm (spec)}_{\rm CT}$ and $\lambda_{\rm Edd}$ for merging U/LIRGs in Figure~\ref{F5-CT-Edd}.
We over-plot the typical relation in Swift/BAT AGNs ($f^{\rm (stat)}_{\rm CT} \sim 0.23$; \citealt{Ricci2017cNature}) and 
\textcolor{black}{mean values of the 52 CT AGN candidates among them ($\sim$0.2--0.5) in the four log$\lambda_{\rm Edd}$ bins,}
the sample of which may have large covering fractions \citep{Tanimoto2022}.
For the late mergers in our sample, which have high ${\log}\lambda_{\rm Edd} \gtrsim -$2, the AGNs that are [\ion{O}{4}] and [\ion{Ne}{5}]-moderate mainly have $f^{\rm (spec)}_{\rm CT}$ close to the value expected from the typical relation of Swift/BAT AGNs and smaller than those of CT AGN candidates, whereas some of the [\ion{O}{4}]-weak and [\ion{Ne}{5}]-weak AGNs show larger $f^{\rm (spec)}_{\rm CT}$ ($\gtrsim$0.4).
A KS test suggests that the difference in $f^{\rm (spec)}_{\rm CT}$ between AGNs showing [\ion{O}{4}] and [\ion{Ne}{5}] weakness and the other AGNs in late mergers is significant at a ${>}90$\% confidence level ($p$-value = 0.074).
Intriguingly, $f^{\rm (spec)}_{\rm CT}$ of some of the sources at close to the effective Eddington limit for CT material (${\log}\lambda_{\rm Edd}^{\rm eff} \sim -$1; e.g., \citealt{Ishibashi2018,Venanzi2020}) is very high, close to $f^{\rm (spec)}_{\rm CT} \sim 1$.

\begin{figure*}
    \centering
    \includegraphics[keepaspectratio, scale=0.58]{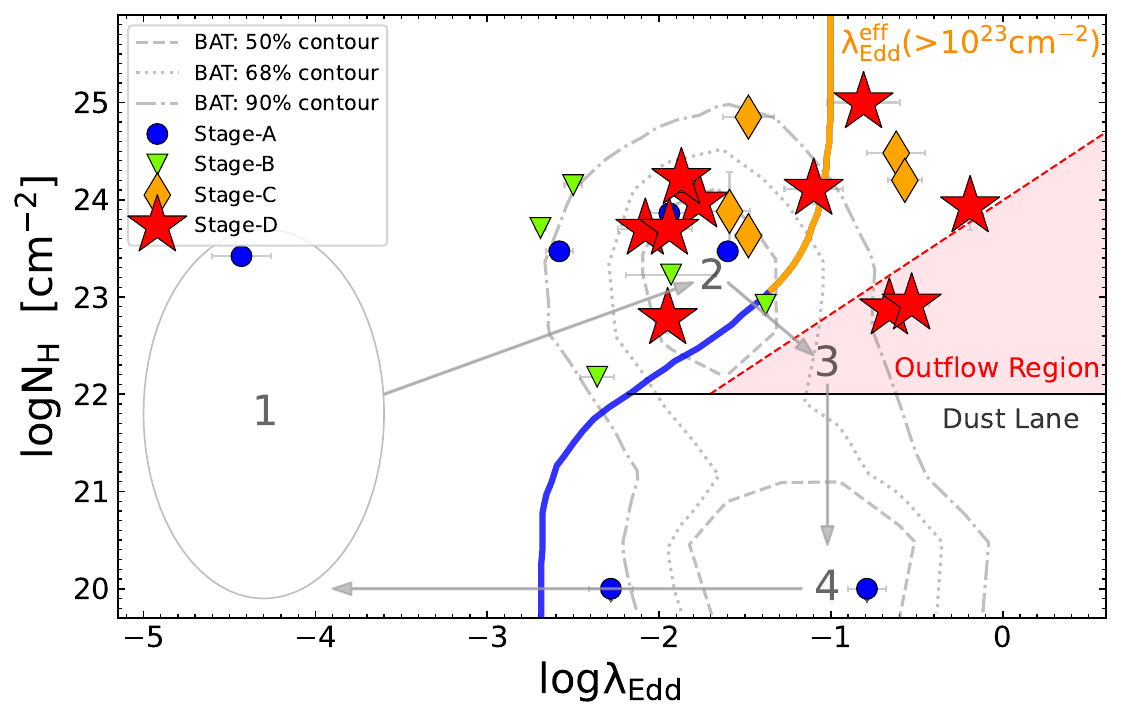}
    \includegraphics[keepaspectratio, scale=0.58]{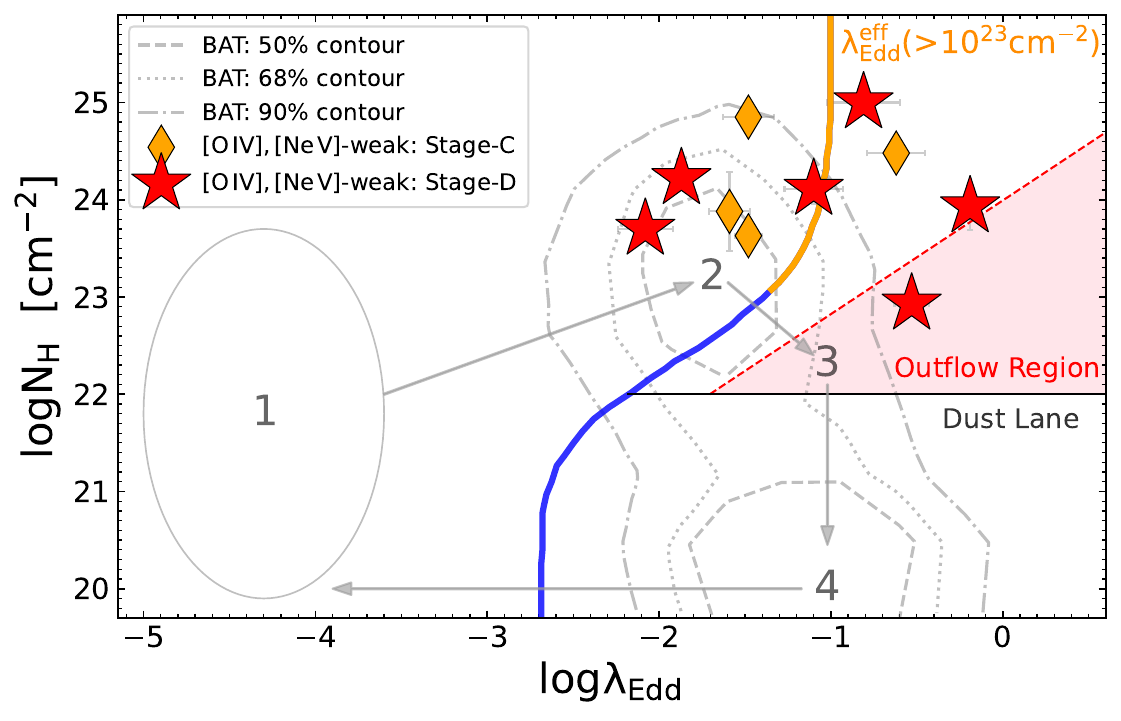}
    \caption{Top panel: $N_{\rm H}$--$\lambda_{\rm Edd}$ diagram \textcolor{black}{for 26 AGNs in U/LIRGs whose Eddington ratios are constrained (Table~\ref{T1-AGN-property}).}
    The (partial) horizontal line at $N_{\rm H} = 10^{22}$~cm$^{-2}$ represents the effective Eddington limit for dusty gas and the pink shaded area does the outflowing region expected according to $\lambda_{\rm Edd}^{\rm eff}$ \citep{Fabian2008}.
    The solid curve is the theoretical evolution path with $\lambda_{\rm Edd}^{\rm eff}$ for obscurers with $N_{\rm H}$ larger (in orange) or smaller (in blue) than $10^{23}$~cm$^{-2}$, in which the IR radiation trapping effect is considered \citep{Ishibashi2018}.
    All gray lines, arrows, and numbers are adopted from \citet{Ricci2022} and show their proposed recurrent evolution process of AGNs; the ellipse labeled as 1 demonstrates a hypothetical region for mostly unobserved AGNs and the three levels of contour lines (in the region with labels of 2, 3, and 4) show 50\%, 68\%, and 90\% of the observed cumulative population among the BAT sources in the ${-}3 < {\log}\lambda_{\rm Edd} < 0$ range (see legend at the top-left and main text for detail). 
    The notations for the other symbols are the same as in Figure~\ref{F1-oiv-nev-12um}.
    \textcolor{black}{Bottom panel: the same as the top but for 10 out of 13 [\ion{O}{4}]-weak or [\ion{Ne}{5}]-weak AGNs in late mergers (Section~\ref{sub3-3-oivweak-fraction}), excluding three ones whose $\lambda_{\rm Edd}$ are not constrained.}
    \\}
\label{F6-NH-Edd}
\end{figure*}

\subsubsection{$N_{\rm H}$--$\lambda_{\rm Edd}$ Diagram}\label{subsub4-1-2-CT-Edd}
\textcolor{black}{The top panel of Figure~\ref{F6-NH-Edd}} plots the $N_{\rm H}$--$\lambda_{\rm Edd}$ diagram in a similar way as in \citet{Fabian2008}. 
In the region below $N_{\rm H} = 10^{22}$~cm$^{-2}$ (solid line) in the diagram, galactic dust lanes may become dominant in absorption. 
Part of the region above the threshold $N_{\rm H} = 10^{22}$~cm$^{-2}$ and at high $\lambda_{\rm Edd}$ of close to 1 or above is the ``outflow'' region (pink-shaded in the figure) where dusty clouds are blown away by radiation pressure.
We also overlay a theoretical AGN evolution-cycle curve adopted from \citet{Ishibashi2018} (blue and orange solid curves). 
We superpose in the diagram the distribution of Swift/BAT AGNs in contours taken from \citet{Ricci2022} and also their proposed radiation-regulated unification model with arrows and number labels, which is concisely described below.
When the AGN is neither very active nor obscured, accretion events enhance mass inflows and obscuration (region 1), and the AGN eventually transits to an obscured AGN (region 2).
Once $\lambda_{\rm Edd}^{\rm eff}$ has increased beyond a certain threshold, depending on the amount of $N_{\rm H}$, which is usually ${\log}N_{\rm H} =$ 22--23 according to the theoretical prediction by \citet{Ishibashi2018} (blue line in the figure) and at most 24, its circumnuclear material is blown away in a short time due to the radiation pressure (region 3).
It soon becomes an unobscured AGN with high $\lambda_{\rm Edd}$ (region 4),
and transits to low $N_{\rm H}$ and $\lambda_{\rm Edd}$.
\textcolor{black}{Most of AGNs in U/LIRGs are distributed on the contours of regions 2--4, while some AGNs in late mergers have larger $N_{\rm H}$ and $\lambda_{\rm Edd}$.}

\textcolor{black}{In the bottom panel of Figure~\ref{F6-NH-Edd}, we make a plot of the same diagram for the [\ion{O}{4}]-weak or [\ion{Ne}{5}]-weak AGNs in late mergers, i.e., buried AGN candidates.
We find that their distribution is mostly concentrated at a higher $N_{\rm H}$--$\lambda_{\rm Edd}$ corner on the $N_{\rm H}$--$\lambda_{\rm Edd}$ plane unlike those for the Swift/BAT AGNs and the other AGNs in U/LIRGs.} 
We also find that the highest $N_{\rm H}$ is $\gtrsim$10$^{25}$~cm$^{-2}$ with $\lambda_{\rm Edd}^{\rm eff}$ close to the effective Eddington limit according to the prediction by \citet{Ishibashi2018} for the AGNs with $N_{\rm H} \geq 10^{23}$~cm$^{-2}$ and CT material (orange curve).
The average of the six buried BAT-IRS AGNs (Section~\ref{sub3-3-oivweak-fraction}) show a similar result, ${\log}N_{\rm H} = 23.9 \pm 0.2$ and ${\log}\lambda_{\rm Edd} = -1.89 \pm 0.28$.
The two AGNs with ${\log}\lambda_{\rm Edd} \sim -$0.5--0 in stage-D mergers in and in the vicinity of the ``outflow'' region, Mrk~231 and IRAS F08572$+$3915, are known to have a kilo-parsec-scale, thus strong, ionized and molecular outflows \citep{Yamada2021}.

The difference in the distributions of Swift/BAT AGNs and buried AGNs on the $N_{\rm H}$--$\lambda_{\rm Edd}$ plane may originate in the difference in the amount of material supplied from the host galaxies.
\textcolor{black}{The circumnuclear material in Swift/BAT AGNs, most of which is nonmergers, are mostly surrounded by obscurers with $N_{\rm H} \geq 10^{22}$~cm$^{-2}$, which would be blown away at ${\log}\lambda_{\rm Edd} \gtrsim -$2 \citep[e.g.,][]{Ricci2022,Ricci2023d},} whereas the circumnuclear material in buried AGNs in late mergers may be shrouded in CT obscurers, which can be blown away at a higher threshold $\lambda_{\rm Edd}$ of $\gtrsim -1$.
In fact, a dramatic variability of $N_{\rm H}$ has been discovered in an AGN with $\lambda_{\rm Edd} \sim -$0.7 (stage-D merger IRAS F05189$-$2524); its column density varied between non-CT ($N_{\rm H} \sim 8 \times 10^{22}$~cm$^{-2}$) and CT (${>}2.3 \times 10^{24}$~cm$^{-2}$) within $\sim$10 years (see Figure~12 in \citealt{Yamada2021}).
This is distinctively larger variability than those in normal AGNs \citep[e.g.,][]{Laha2020}.
Thus, these facts support that a larger amount of CT material in buried AGNs than in Swift/BAT AGNs causes a unique structure of buried AGNs in late mergers, and hence, the variety of circumnuclear material may be uniformly regulated by the radiation pressure depending on $\lambda_{\rm Edd}$ and the amount of material supplied, which is a key parameter of the difference between mergers and nonmergers.

\subsection{Future Observations with X-ray and IR Spectroscopy}
A combination of deep X-ray and mid-IR spectroscopy would be crucial to reveal the fraction and structure of buried AGNs in U/LIRGs, which are the key population in the study of SMBH growth in obscured AGNs at $z \sim 1$ and beyond (see Section~\ref{S1-intro}).
James Webb Space Telescope (JWST), endowed with a supreme mid-IR spectroscopic capability and unprecedentedly high sensitivity, can identify rapidly growing SMBHs in U/LIRGs \citep[e.g.,][]{Inami2022b,Rich2023}.
JWST observations combined with high-quality X-ray observations would enable us to identify buried AGNs almost fully covered by dense circumnuclear material with high $f^{\rm (spec)}_{\rm CT}$ at cosmic noon, the population of which is predicted to increase significantly toward higher redshifts \citep[e.g.,][]{Ueda2014,Gilli2022}, and would help us reveal the entire picture of growth of both obscured and unobscured SMBHs.
\\

\section{Conclusions}\label{S5-conclusion}
\textcolor{black}{We established new mid-IR diagnostics employing $L_{\rm [O\,IV]}$/$L_{\rm 12,AGN}$ or $L_{\rm [Ne\,V]}$/$L_{\rm 12,AGN}$ for the buried AGN almost fully covered by dense circumnuclear material.
We evaluated the method by comparing it with the estimates of covering fractions derived from the X-ray spectral fitting with XCLUMPY.
This study identified 17 buried AGN candidates in merging U/LIRGs and 6 ones from nonmerging BAT-IRS sources.}

Constructing the cumulative $N_{\rm H}$ distribution, we estimated the covering fraction of CT obscurers $f^{\rm (spec)}_{\rm CT} \sim 0.5 \pm 0.1$ for the buried AGNs in U/LIRGs, the result of which is consistent with the estimates with XCLUMPY. 
The $f^{\rm (spec)}_{\rm CT}$ value is much larger than that for Swift/BAT AGNs ($f^{\rm (stat)}_{\rm CT} = 23 \pm 6$\%; \citealt{Ricci2017cNature}).
\textcolor{black}{Their large fraction of obscured AGNs with $N_{\rm H} \geq 10^{22}$~cm$^{-2}$ (${\gtrsim}90$\%) supports that the [\ion{O}{4}]-weak and [\ion{Ne}{5}]-weak AGNs are likely to be ``buried'', with the AGN not being able to develop a NLR due to optically-thick material in almost all directions.}
The fraction of [\ion{O}{4}]-weak AGNs increases with merger stage from $27^{+13}_{-10}$\% (early) to $66^{+10}_{-12}$\% (late mergers).
Similar results are obtained with the [\ion{Ne}{5}] line.
For the BAT-IRS AGNs, the fraction of [\ion{O}{4}] or [\ion{Ne}{5}]-weak AGNs with $N_{\rm H} \geq 10^{23}$~cm$^{-2}$, plausible buried AGN candidates, is only $5 \pm 2$\% (6/138).
Considering that the fraction is much smaller than that of buried AGNs in late mergers ($66^{+10}_{-12}$\%), most of which are obscured AGNs (${\gtrsim}90\%$), galaxy merging is likely to enhance the buried structure.

We also found that the [\ion{O}{4}]-weak or [\ion{Ne}{5}]-weak AGNs in late mergers show larger $N_{\rm H}$ and $\lambda_{\rm Edd}$ than those of the Swift/BAT AGNs (Figure~\ref{F6-NH-Edd}), and the largest $N_{\rm H}$ is $\gtrsim$10$^{25}$~cm$^{-2}$ with ${\log}\lambda_{\rm Edd} \sim -$1, close to the effective Eddington limit for CT material.
These results suggest that (1) the circumnuclear material in buried AGNs is regulated by the radiation pressure from high-$\lambda_{\rm Edd}$ AGNs on the CT obscurers, and 
(2) their dense material with large $f^{\rm (spec)}_{\rm CT}$ ($\sim 0.5 \pm 0.1$) in U/LIRGs is a likely cause of a unique structure of this type of buried AGNs and the amount of material may be maintained through merger-induced continuous supply from their host galaxies.
Exploiting the [\ion{O}{4}] and [\ion{Ne}{5}] weakness, buried AGNs at higher-$z$, which may be a majority of the AGNs at $z \gtrsim 1$, can be identified with a combination of X-ray and mid-IR spectroscopy, particularly in the era of astronomy with JWST.
\\

\textcolor{black}{The authors thank the anonymous referee for very valuable comments and suggestions which improved the manuscript.}
This work is supported by JSPS KAKENHI grant numbers 
22K20391 and 23K13154 (S.Y.), 
20H01946 (Y.U.), 
23K13153 (T.K.),
19K14759 and 22H01266 (Y.T.), 
21K03632 (M.I.),
20H01939 (K.I.),
22J22795 (R.U.),
and 21H04496 (K.W.).
S.Y. and T.K. are grateful for support from the RIKEN Special Postdoctoral Researcher Program.
C.R. is grateful for support from the Fondecyt Regular grant 1230345 and ANID BASAL project FB210003.
T.M. and M.H.E. acknowledge support by UNAM-DGAPA PAPIIT IN114423 
and CONACyT Investigaci\'on Científica B\'asica 252531.
A.T. is supported by the Kagoshima University postdoctoral research program (KU-DREAM).
H.M.E. is supported by a postdoctoral fellowship from UNAM-DGAPA.


\begin{deluxetable*}{lccccccc}
\label{T1-AGN-property}
\tablecaption{Main Properties of Host Galaxies and AGNs in local U/LIRGs and Swift/BAT Sources}
\tabletypesize{\footnotesize}
\tablehead{
\colhead{Object} &
\colhead{$M$/id} &
\colhead{$z$} &
\colhead{$D_{\rm L}$} &
\colhead{${\log}N_{\rm H}$} &
\colhead{$f^{\rm (spec)}_{\rm CT}$} &
\colhead{${\log}L_{\rm AGN}$} &
\colhead{${\log}M_{\rm BH}$} \\
\ \ \ \ \ \ \ \ (1) & (2) & (3) & (4) & (5) & (6) & (7) & (8) \\
 & ${\log}\lambda_{\rm Edd}$ & ${\log}L_{\rm 12,AGN}$ &
${\log}L_{\rm [O\,IV]}$ & ${\log}L_{\rm [Ne\,V]}$ & ${\log}R_{\rm [O\,IV]}$ & ${\log}R_{\rm [Ne\,V]}$ & O/N-Weak \\
& (9) & (10) & (11) & (12) & (13) & (14) & (15)
}
\startdata
\multicolumn{8}{c}{Sample 1: AGNs in U/LIRGs \textcolor{black}{(30 objects)}} \\
\hline
NGC~833 & A & 0.01290 & 55.8 & $23.42^{+0.01}_{-0.01}$& $0.43^{+0.01}_{-0.02}$ & $42.28 \pm 0.17$ & 8.61 \\ & $-4.43 \pm 0.17$ & 41.95 & ${<}39.34$ & ${<}39.60$ & ${<}{-}2.61$ & ${<}{-}2.35$ & n/n \\ \hline
\hline
\multicolumn{8}{c}{Sample 2: Swift/BAT AGNs in \citet{Ogawa2021} \textcolor{black}{(14 objects)}} \\
\hline
NGC~2110 & 308 & 0.0075 & 34.3 & $22.60^{+0.06}_{-0.08}$& ${<}0.01$ & 44.50 & 8.78 \\ & $-2.46$ & 42.89 & $40.80 \pm 0.01$ & $40.06 \pm 0.08$ & $-2.09 \pm 0.01$ & $-2.84 \pm 0.08$ & n/n \\ \hline
\hline
\multicolumn{8}{c}{Sample 3: [\ion{O}{4}] or [\ion{Ne}{5}]-weak obscured AGNs in BAT-IRS sources \textcolor{black}{(10 objects)}} \\
\hline
ESO~297$-$18 & 87 & 0.0252 & 109.8 & $23.83^{+0.07}_{-0.07}$& $\cdots$ & 45.00 & 8.50 \\ & $-1.67$ & 43.24 & ${<}40.33$ & $40.13 \pm 0.08$ & ${<}{-}2.91$ & $-3.11 \pm 0.08$ & Y/n \\ \hline
\enddata
\tablecomments{Columns:
(1) object name;
(2) merger stage (U/LIRGs) or ID in Swift/BAT 70-month-observation catalog;
(3)--(4) redshift and luminosity distance.
We utilize the distance for NGC 4418 (34 Mpc; \citealt{Yamada2021}) and close objects at $<$50~Mpc in Swift/BAT AGNs from \citet{Koss2022b};
(5) logarithmic $N_{\rm H}$ in cm$^{-2}$;
(6) covering fraction of CT material. 
For the faint AGNs whose torus angular width are fixed at $20\degr$, we calculate the value by adding a typical uncertainty ($20 \pm 5\degr$; e.g., \citealt{Yamada2021});
(7)--(9) logarithmic bolometric AGN luminosity, SMBH mass, and Eddington ratio.
(10)--(12) 12-$\mu$m AGN luminosity, [\ion{O}{4}] luminosity, and [\ion{Ne}{5}] luminosity;
(13)--(14) logarithmic ratios of mid-IR line luminosity to 12-$\mu$m AGN luminosity for [\ion{O}{4}] ($R_{\rm [O\,IV]}$) and [\ion{Ne}{5}] ($R_{\rm [Ne\,V]}$).
(15) criteria of buried AGN candidates based on the $R_{\rm [O\,IV]}$ and $R_{\rm [Ne\,V]}$ (see Section~\ref{sub3-1-oiv-weak}). Y mean the [\ion{O}{4}]-weak (left) or [\ion{Ne}{5}]-weak AGNs (right), while n means [\ion{O}{4}]-moderate (left) or [\ion{Ne}{5}]-moderate AGNs (right).
The references are summarized in Section~\ref{S2-data} and Appendix~\ref{AppendixB-BAT}.
}
\tablenotetext{\ }{(This table is available in its entirety in machine-readable form.)}
\end{deluxetable*}

\appendix
\restartappendixnumbering

\section{Detail of GOALS AGN Candidate Evaluation} 
\subsection{SED-identified three AGN candidates} \label{AppendixA1-CTAGN}
\citet{Yamada2023} identified, on the basis of SED decomposition, three AGN candidates 
(ESO 350$-$38, ESO 374$-$032, and NGC~4418), which had been observed in X-rays, specifically with NuSTAR in the 8--24~keV band, but not been detected presumably due to insufficient exposures ($\lesssim$50~ks) (see Section~\ref{sub2-1-GOALS}).
Their WISE color criteria \citep{Stern2012} and equivalent widths of the 6.2~$\mu$m polycyclic aromatic hydrocarbon features suggest that they are obscured AGNs \citep[e.g.,][]{Yamada2021}.
Using Equation~7 in \citet{Asmus2015} in combination with their 12-$\mu$m AGN fluxes and observed 2--10~keV fluxes, we estimate their ${\log}N_{\rm H}$ to be $25.3\pm1.8$ (ESO 350$-$38), $25.6\pm1.9$ (ESO 374$-$032), and $26.2\pm1.9$ (NGC~4418).
The 3$\sigma$ upper limits of intrinsic 2--10~keV luminosities are then estimated to be ${\log}L_{2-10}^{\rm (lim)} = 42.42$, 42.69, and 41.49, respectively, from the 3$\sigma$ upper limit of their 8--24~keV luminosities under the assumption of ${\log}N_{\rm H} = 25$ \citep{Yamada2021}. 
Comparison between their X-ray and bolometric AGN luminosities yields 2--10~keV bolometric correction factors of 20--160. 
The values are roughly consistent with the typical one among local AGNs with a wide range of ${\log}\lambda_{\rm Edd} = -$3--0 \citep{Vasudevan2007}.
Thus, these three sources are expected to be heavily obscured AGNs with ${\log}N_{\rm H} \gtrsim 25$, or pure starbursts.

\subsection{\textcolor{black}{Seven AGNs in Nonmerging U/LIRGs}} \label{AppendixA2-nonmerger}
\textcolor{black}{To discuss the difference between merging U/LIRGs (Section~\ref{sub2-1-GOALS}) and nonmerging BAT-IRS AGNs (Section~\ref{sub2-2-BAT-AGN}), seven nonmergers in GOALS sample are excluded from the U/LIRG sample.
Among them, NGC~1068, NGC~1275, and NGC~1365 are the AGNs with complex spectral features in the X-ray band, whose circumnuclear structures are difficult to be constrained \citep{Yamada2021}.
For the other AGNs (UGC~2608, MCG$-$03-34-064, NGC 5135, and NGC 7130), they are obscured with ${\sim}10^{23}$--$10^{25}$~cm$^{-2}$ and have
$f^{\rm (spec)}_{\rm CT} \sim 0.2$--0.5, consistent with the typical values of CT AGNs in Swift/BAT AGNs \citep{Tanimoto2022}.
These AGNs are neither [\ion{O}{4}]-weak nor [\ion{Ne}{5}]-weak AGNs, unlike the AGNs in late mergers \citep[see also][]{Yamada2020}.}
\\

\section{Constructing a BAT-IRS AGN sample set} \label{AppendixB-BAT}
This section describes the procedure and criteria to construct the BAT-IRS AGN sample set (Section~\ref{sub2-2-BAT-AGN}) used in this work.
\textcolor{black}{The parent sample set of X-ray selected AGNs for our X-ray AGN sample set is 606 non-blazer AGNs in the Swift/BAT 70-month-observation catalog at galactic latitudes ($|b| > 10$\degr) whose spectroscopic redshifts are available \citep{Ichikawa2017}.
Decomposing the 3--500~$\mu$m SEDs, \citet{Ichikawa2019a} estimated the AGN contribution to the IR luminosities for 587 sources, where the interacting galaxies not spatially resolved with the Swift/BAT are excluded.
From them, we also exclude the interacting galaxies in GOALS sample \citep{Yamada2021} and three dual-AGN systems in \citet{Koss2012}; i.e., Mrk~463 \citep{Bianchi2008,Yamada2018}, NGC~3227, and NGC~2992 \citep{Ricci2017dApJS}.
Among them, both [\ion{O}{4}] and [\ion{Ne}{5}] line fluxes of 143 sources are measured (n.b., only upper limits are determined for some of them) by Spitzer/IRS spectra, which were obtained from IDEOS\footnote{IDEOS (Infrared Database of Extragalactic Observables from Spitzer, catalog) catalog builds on the publication-quality Spitzer/IRS spectra, containing 3558 sources out of 5015 sources in the CASSIS repository (\url{https://cassis.sirtf.com}).} catalog \citep{Spoon2022}.
We select the AGNs whose 12~$\mu$m AGN luminosities \citep{Ichikawa2019a}, bolometric AGN luminosities derived from intrinsic 14--150~keV ones, SMBH masses, and Eddington ratios \citep{Koss2022b} are fully determined.
We adopt the redshifts and luminosity distances in \citet{Koss2022b}, in which redshift-independent distances are presented for close objects at $<$50~Mpc.
The resultant sample totals 138 BAT-IRS AGNs, which are used in this work.}
\\

\section{Definition of Buried AGNs for Poorly-Constrained Data Points} \label{AppendixC-uplim}
In Section~\ref{sub3-1-oiv-weak}, we have adopted the [\ion{O}{4}]-weak thresholds of ${\log}L_{\rm [O\,IV]}$/$L_{\rm 12,AGN} \leq -$3.0 or $-2.7$, depending on whether the data points are well determined or have upper limits only. 
Here we consider the case of universal application of the former threshold to both types of data points in our sample.
\textcolor{black}{In this case, the covering fractions of [\ion{O}{4}]-weak AGNs are 
$f^{\rm (stat)}_{\rm CT} = 46 \pm 14\%$ (5/11) and 
$f^{\rm (stat)}_{\rm obs}({\geq}10^{22}~{\rm cm}^{-2}) = 94^{+5}_{-8}\%$ (11/11).}
Similarly, universal application of the threshold of ${\log}L_{\rm [Ne\,V]}$/$L_{\rm 12,AGN} \leq -$3.4 
to our sample yields fractions of 
$f^{\rm (stat)}_{\rm CT} = 50 \pm 12\%$ (7/14) and 
$f^{\rm (stat)}_{\rm obs}({\geq}10^{22}~{\rm cm}^{-2}) = 89^{+6}_{-9}\%$ (13/14).
Their average $f^{\rm (spec)}_{\rm CT}$ estimated with XCLUMPY are $0.58\pm0.19$ for [\ion{O}{4}]-weak AGNs 
and $0.51\pm0.22$ for [\ion{Ne}{5}]-weak ones, which are consistent with the CT AGN fractions.
In consequence, if these thresholds are adopted, $f^{\rm (spec)}_{\rm CT}$ for buried AGNs are typically $f^{\rm (spec)}_{\rm CT} \sim 0.5 \pm 0.1$.


\bibliography{sample631}{}
\bibliographystyle{aasjournal}



\end{document}